\documentclass[12pt]{article}
\usepackage{amsmath}

\usepackage{amssymb}
\usepackage{amsmath,axodraw}
\usepackage{euscript}

\usepackage{epsfig}  
  
\usepackage{cite}  
  
\usepackage{alltt}  
\def \bea{\begin{eqnarray}}
\def \beq{\begin{equation}}

\def \to {\rightarrow}
\def \eea{\end{eqnarray}}
\def \eeq{\end{equation}}

\begin{document}
\begin{titlepage}
%Finding the CP Nature of the Higgs Boson at the NLC.
\begin{flushright}  
{\bf   
IFT-P.0xx/2002 \\
July 2002}  
\end{flushright}  

\hspace{1 mm}  
\vspace{1 cm}  
\begin{center}{\bf\Large { Triple Photon Production at the Tevatron in
Technicolor Models.}}
\end{center}  
%\begin{center}{\bf\Large  Tevatron and its Three-Photon Signature.} \end{center}   
\vspace{0.3 cm}  
\begin{center}  
{\large\bf A. Zerwekh$^{a,}$ \footnote{alfonso.zerwekh@fis.utfsm.cl} ,~ C.
Dib$^{a,}$ \footnote{cdib@fis.utfsm.cl}  ~ and ~  R. Rosenfeld$^{b,}$
\footnote{rosenfel@ift.unesp.br}}  
\vspace{0.3 cm}  
\\  
{\em $^a$ Department of Physics, Universidad T\'ecnica Federico Santa Mar\'{\i}a \\
Valpara\'{\i}so, Chile}\\  
{\em $^b$ Instituto de F\'{\i}sica Te\'orica - UNESP \\
 Rua Pamplona, 145, 01405-900 S\~ao Paulo, SP, Brasil}\\    
\vspace{0.5 cm}  
  
\end{center}

\centerline {\bf Abstract}
\vskip 1.0cm

We study the process $p \bar{p} \to \gamma \gamma \gamma$ as a signal for
associated photon-technipion production at the Tevatron. This is a clean
signature with relatively low background. Resonant and
non-resonant contributions are included and we show that technicolor models can
be effectively probed in this mode. 

\bigskip

\vfill
\end{titlepage}

\newpage

\section{Introduction}

\indent
The origin of fermion masses and mixings is one of most important issues in
particle physics. Unfortunately, these parameters are inputs in the well-tested 
Standard Model (SM). 
Fermion masses are possibly related to the electroweak symmetry 
breaking mechanism, which is not known at the
moment and is the top priority of present and future experiments. 
In the SM, a scalar electroweak doublet with self-interactions described by 
an {\it ad hoc} quartic
potential is responsible for the symmetry breaking, leaving a scalar physical
boson, the Higgs boson (J$^{\mbox{{\tiny PC}}} = 0^{++}$), as a remnant. 
Favorite extensions of the SM, like  the minimal supersymmetric standard model
(MSSM) \cite{mssm}, also predict the existence of a heavy pseudoscalar boson 
(J$^{\mbox{{\tiny PC}}} = 0^{-+}$), in addition to a 
light scalar boson. 

Another interesting possibility is that the electroweak
symmetry breaking is triggered by some new strong interaction, generally called
technicolor, and in this case
the lightest boson could be a pseudoscalar, 
like a pion, named technipion. In fact, in these models of
dynamical symmetry breaking a whole new set of resonances related to the 
technicolor sector is predicted \cite{techni}.  

It is important to find experimental signatures that can distinguish these 
different models of symmetry breaking. A compilation of  experimental
signatures for different technicolor models, like multiscale  and
top-color assisted walking technicolor, can be found in \cite{signatures}.

In this letter we focus on the signature arising from associated
photon-technipion production. This is analogous to the associated gauge-higgs 
boson production. The process 
$p \bar{p} \to \Pi_T^{(\prime)} (\gamma, Z, W^{\pm})$, 
where $\Pi_T^{(\prime)}$ is a  isospin triplet (singlet) technipion,  
can be enhanced by low-lying
technicolor resonances like the techni-rho and the techni-omega. These processes
have been studied in \cite{associated} and the importance of the process
involving the final state photon has been stressed in \cite{lane0}.  

The process $e^+ e^- \to \gamma \Pi_T^{(\prime)}$ 
was analysed for LEP and future linear 
colliders in \cite{lp}. Recently, Lane {\it et al.} \cite{lane1} re-studied this 
process taking into account both continuum and resonance contributions,  
but concentrating on the dominant $b \bar{b}$ decay mode.

In this letter we study the possibility of using the process
$p \bar{p} \to \gamma \Pi_T^{(\prime)} \to \gamma \gamma \gamma$, which is a 
clean
signature with relatively low background even in a hadronic environment, to 
put some constraints in some
technicolor models. We also include both resonant and non-resonant contributions
in our analysis and perform a simulation of the significance level of this
signature. 

\section{The Model}

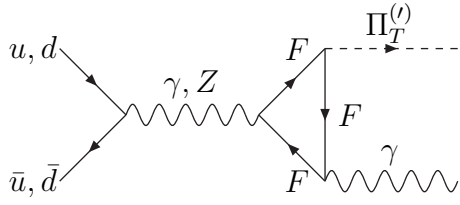
\begin{figure}[htbp]
  \begin{center}
    \begin{picture}(400,100)(0,0)
      \ArrowLine(50,75)(75,50)
      \Text(40,75)[]{$u,d$}
      \ArrowLine(75,50)(50,25)
      \Text(40,25)[]{$\bar{u},\bar{d}$}
      \Photon(75,50)(125,50){4}{5}
      \Text(100,62)[]{$\gamma,Z$}
      \ArrowLine(125,50)(150,75)
      \Text(140,75)[]{$F$}
      \ArrowLine(150,25)(125,50)
      \ArrowLine(150,75)(150,25)
      \Text(140,25)[]{$F$}
      \Text(160,50)[]{$F$}
      \DashArrowLine(150,75)(200,75){3}
      \Text(175,85)[]{$\Pi_T^{(\prime)}$}
      \Photon(150,25)(200,25){4}{5}
      \Text(175,35)[]{$\gamma$}
    \end{picture}
    \caption{Triangle anomaly giving the continuum contribution to the $p
    \bar{p} \to \Pi_T^{(\prime)} \gamma$ process.  }
    \label{fig:triangle}
  \end{center}
\end{figure}

The coupling of the technipion to two gauge bosons is mediated by the
Adler-Bell-Jackiw anomaly \cite{abj} arising from a techniquark triangle. 
The $\Pi_T^{(\prime)} B_1 B_2$ coupling can be
parametrized as:

\begin{equation}
A_{\Pi_T B_1 B_2} = \frac{ S_{\Pi_T B_1 B_2}}{4 \sqrt{2} \pi^2 F_{\Pi_T}}
\epsilon_{\mu\nu\alpha\beta} \varepsilon_1^\mu \varepsilon_2^\nu k_1^\alpha
k_2^\beta ,
\end{equation}
where $\varepsilon_{1,2}$ and $k_{1,2}$ are the polarization vectors and momenta
of the gauge bosons $B_{1,2}$ respectively. $F_{\Pi_T}$ is the technipion decay
constant, which is related to the technipion coupling to the axial current. 
The group-theoretical factor 
$S_{\Pi_T B_1 B_2}$ is given by \cite{ellis}:

\begin{equation}
S_{\Pi_T B_1 B_2} = g_1 g_2 \; Tr \left( Q_{\Pi_T} 
\frac{\lbrace Q_1,Q_2 \rbrace}{2} \right)
\end{equation}
where $g_1$ and $g_2$ are the corresponding gauge coupling constants and $Q_1$,
$Q_2$ and $ Q_{\Pi_T}$ are the charges under the gauge groups and 
isospin respectively of the technifermions circulating in the loop. 
For our purposes
we will be concerned only with the $\Pi_T^{(\prime)} \gamma \gamma$ and 
$\Pi_T^{(\prime)}  \gamma Z$
couplings, since they provide the only contributions to the process
$p \bar{p} \to \gamma \Pi_T^{(\prime)}$, shown in Figure \ref{fig:triangle}, 
and the
corresponding group-theoretical factors, for a one-family
technicolor model with gauge group $SU(N_{TC})$, are given by :

\begin{equation}
S_{\Pi_T \gamma \gamma} = \frac{4 e^2}{ \sqrt{6}} N_{TC} \;\; , \;\;\;\; 
S_{\Pi_T Z \gamma} = 2 e^2 \frac{1 - 4\sin^2 \theta_w }{\sqrt{6} 
\sin 2\theta_w} 
N_{TC}
\end{equation}

\begin{equation}
S_{\Pi_T^\prime \gamma \gamma} = - \frac{4 e^2}{3 \sqrt{6}} N_{TC} \;\; , \;\;\;\; 
S_{\Pi_T^\prime Z \gamma} = \frac{4 e^2 \tan \theta_w}{3 \sqrt{6}} N_{TC}  
\end{equation}

\begin{figure}[htbp]
  \begin{center}
    \begin{picture}(400,100)(0,0)
    \ArrowLine(100,75)(150,50)
    \ArrowLine(150,50)(100,25)
    \Photon(150,50)(200,50){4}{5}
    \SetWidth{5}
    \Line(200,50)(250,50)
    \SetWidth{0.5}
    \DashLine(250,50)(300,75){4}
    \Photon(250,50)(300,25){4}{5}
     \Text(125,75)[]{$u,d$}
     \Text(125,25)[]{$\bar{u},\bar{d}$}
     \Text(175,60)[]{$\gamma,Z$}
     \Text(225,60)[]{$\omega_T,\rho_T$}
     \Text(320,75)[]{$\Pi_T^{(\prime)}$}
     \Text(320,25)[]{$\gamma$}
    \end{picture}
    \caption{Techni-omega and techni-rho contributions to the 
    $p\bar{p} \to \Pi_T^{(\prime)}\gamma$ process.}
    \label{fig:techniomega}
  \end{center}
\end{figure}
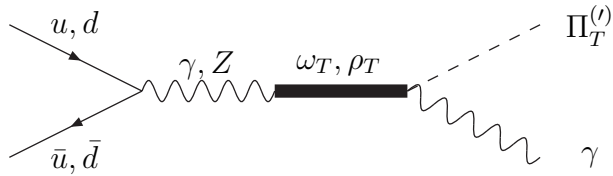

Consequently, the decay of neutral techni-pions into two photons is induced
entirely by the anomaly. In contrast, the associated production of a photon with
a neutral techni-pion is mediated by both $\Pi_T^{(\prime)} \gamma \gamma$ and
$\Pi_T^{(\prime)} \gamma Z$ anomalous vertices, as well as by possible 
$s-$channel vector resonances, the isosinglet techni-omega ($\omega_T$), and the
isotriplet techi-rho ($\rho_T$). These further contributions are depicted in 
Figure \ref{fig:techniomega} and can be treated as a
generalization of vector meson
dominance. 

From the viewpoint of perturbation theory, the anomalous couplings appear only
at the one-loop level. The resonances, considered as techniquark bound states,
are a sum to all orders in technicolor interactions and therefore include 
one-loop effects. However, no ambiguity of double-counting arises when we 
consider both the anomaly and resonance contributions, as these are due to very
different energy scales, the former being a low-energy effect and therefore is 
not important at the resonance mass scale.

In the absence of isospin violation, the techni-omega mixes with the 
isoscalar part of
the electroweak current, the $B_\mu$ field, whereas the techni-rho mixes with
the isotriplet part, the $W^3_\mu$ field. In terms of the physical fields of the
photon and the Z-boson, the mixing strengths are given by:
\begin{equation}
g_{\omega_T-\gamma} =  \sqrt{\frac{\alpha}{\alpha_T}} \left(Q_U+Q_D\right)    
\;\;,\;\;\;\;
g_{\omega_T-Z} = -\sqrt{\frac{\alpha}{\alpha_T}} \left(Q_U+Q_D\right) \tan
\theta_W
\end{equation}
and
\begin{equation}
g_{\rho_T-\gamma} = \sqrt{\frac{\alpha}{\alpha_T}}       \;\;,\;\;\;\;
g_{\rho_T-Z} = \sqrt{\frac{\alpha}{\alpha_T}} \cot 2 \theta_W
\end{equation}
where $\alpha$ is the fine structure constant and $\alpha_T$ is related to the
technicolor coupling constant $g_T$ and can be estimated by a na\"{\i}ve scaling
from QCD:
\begin{equation}
\alpha_T = \frac{g_T^2}{4 \pi} = 2.9 \; \left(\frac{3}{N_{TC}}\right)
\end{equation}

Finally, the relevant amplitudes for the decays $\rho_T,\; \omega_T \to
\gamma \Pi_T^{(\prime)}$ are given by, in the notation of \cite{lane0}:
\begin{equation}
{\cal M} \left( V_T (q) \to G (p_1) \Pi_T^{(\prime)} (p_2) \right) = 
\frac{e V_{V_T G \Pi_T}}{M_V} 
\epsilon_{\mu\nu\alpha\beta} \varepsilon^\mu(q) \varepsilon^{\ast \nu}(p_1)
q^\alpha p_1^\beta 
\end{equation}
where $M_V$ is a mass parameter usually taken to be $200$ GeV and
\begin{equation}
V_{\omega_T \gamma \Pi_T} =  \cos \chi 
,\;\;
V_{\omega_T \gamma \Pi_T^\prime} = \left(Q_U+Q_D\right) \cos \chi'  
,\;\;
V_{\rho_T \gamma \Pi_T} = \left(Q_U+Q_D\right) \cos \chi
,\;\;
V_{\rho_T \gamma \Pi_T^\prime} = \cos \chi'.
\end{equation}
In the equation above $\chi$ and $\chi'$ are mixing angles between the isospin
eigentates and the mass eigenstates. In our computations we use a value of
$\sin\chi = \sin\chi' = 1/3$ and $ Q_U + Q_D= 5/3 $ \cite{lane0}. 
In order to compute the fermionic widths of the techni-pions
we use the coupling constant $g_{\Pi_T f \bar{f}} = m_f/F_{\Pi_T}$.

%%%%%%%%%%%%%%%%%%%%%%%%%%%%%%%%%%%%%%%%%%%%%%%%%%%
\section{Simulation of the process }

\noindent
The inputs to our codes are the relevant masses of $\Pi_T^{(\prime)}$,
$\omega_T$, $\rho_T$, the technipion decay constant $F_{\Pi_T}$ and the
resonance widths $\Gamma_{\rho_T}$ and $\Gamma_{\omega_T}$. 
In order to reduce the number of parameters, we will use in our calculations 
the reference
set of values $m_{\Pi_T^{(\prime)}} = 110$ GeV and 
$m_{\omega_T} = m_{\rho_T}$. We also adopt $N_{TC}=4$ and 
$F_{\Pi_T} = 82$ GeV, as appropriate
in multiscale walking technicolor, but the results are relatively insensitive to
this choice since the couplings of the vector resonances and the branching ratio
$BR(\Pi_T^{(\prime)} \to \gamma \gamma)$ are independent of $F_{\Pi_T}$.  
The vector resonance widths were obtained from 
Pythia  version 6.125 \cite{pythia}. 

We used the parton distribution function CTEQ6 \cite{cteq} with both momentum and 
factorization scales set at $\sqrt{\hat{s}}$ 
and a total center-of-mass energy of
$\sqrt{s}=2000$ GeV. We convoluted the relevant parton distribution functions
with the amplitudes described above. Total luminosities of $2$ fb$^{-1}$ (Run
2a) and 
$30$ fb$^{-1}$ (optimistic Run 2b) were considered.

The irreducible background was generated using the program CalcHEP 2.1
\cite{calchep}. The main irreducible contribution comes from 
$ u \bar{u}, d \bar{d} \to  \gamma \gamma \gamma $. 

A gaussian smearing for the final state photon energy with 
$\sigma_E/E = 0.20/\sqrt{E}$ \cite{higgs} 
was applied to both signal and background.

\section{Results}

\begin{figure}
\centerline{\epsfxsize=1.5\hsize \epsffile{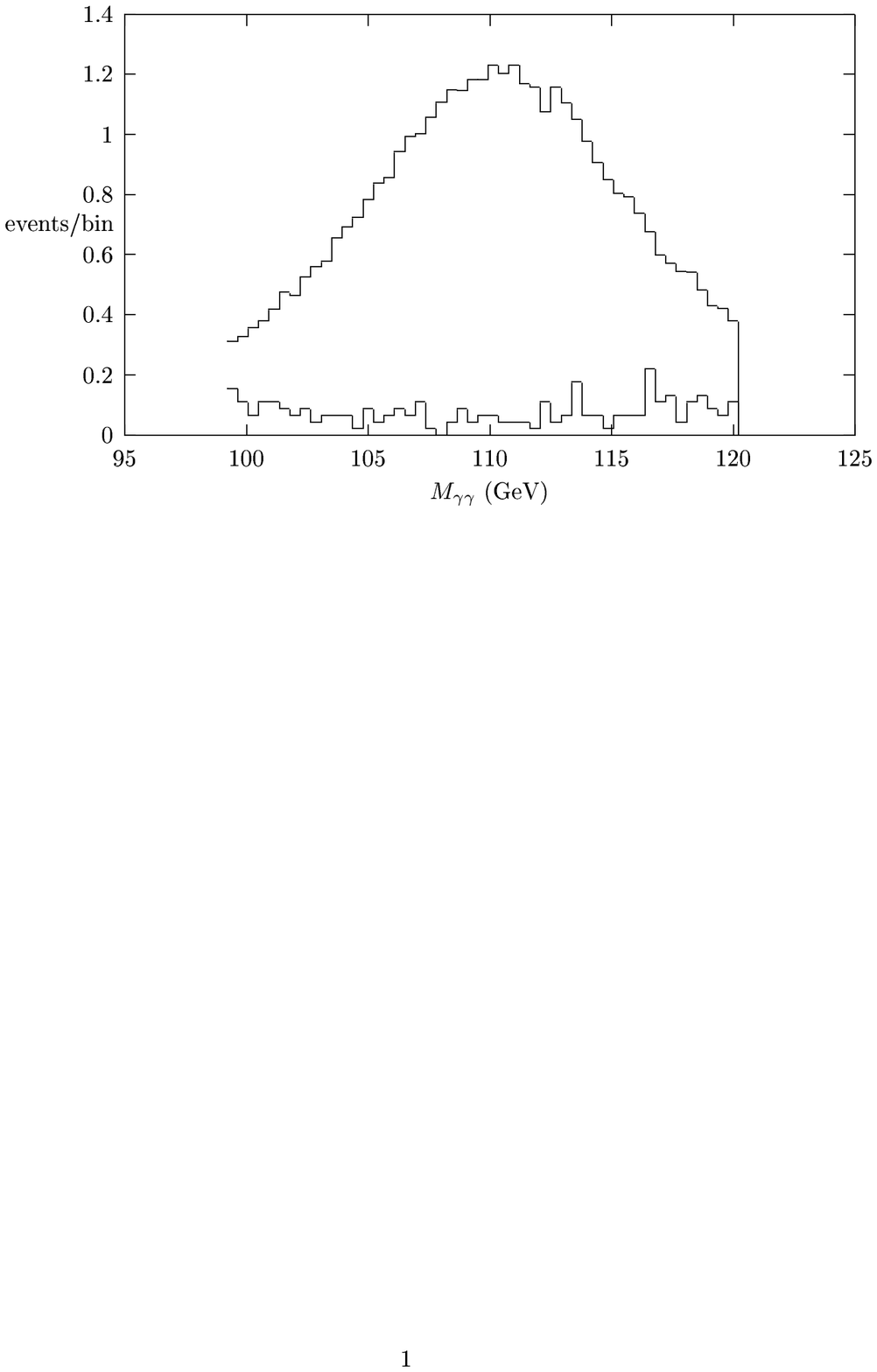}}
\vskip-20cm
\caption{
2-photon invariant mass distribution for signal (upper histogram) and
background (lower histogram) for $m_{\omega_T} = m_{\rho_T} = 350$
GeV and $m_{\Pi_T^{(\prime)}} = 110$ GeV for ${\cal L} = 30$ fb$^{-1}$. The bin
size used in  these histograms is $0.43$ GeV. }
\label{2gamma}
\end{figure}

\begin{figure}
\centerline{\epsfxsize=1.5\hsize \epsffile{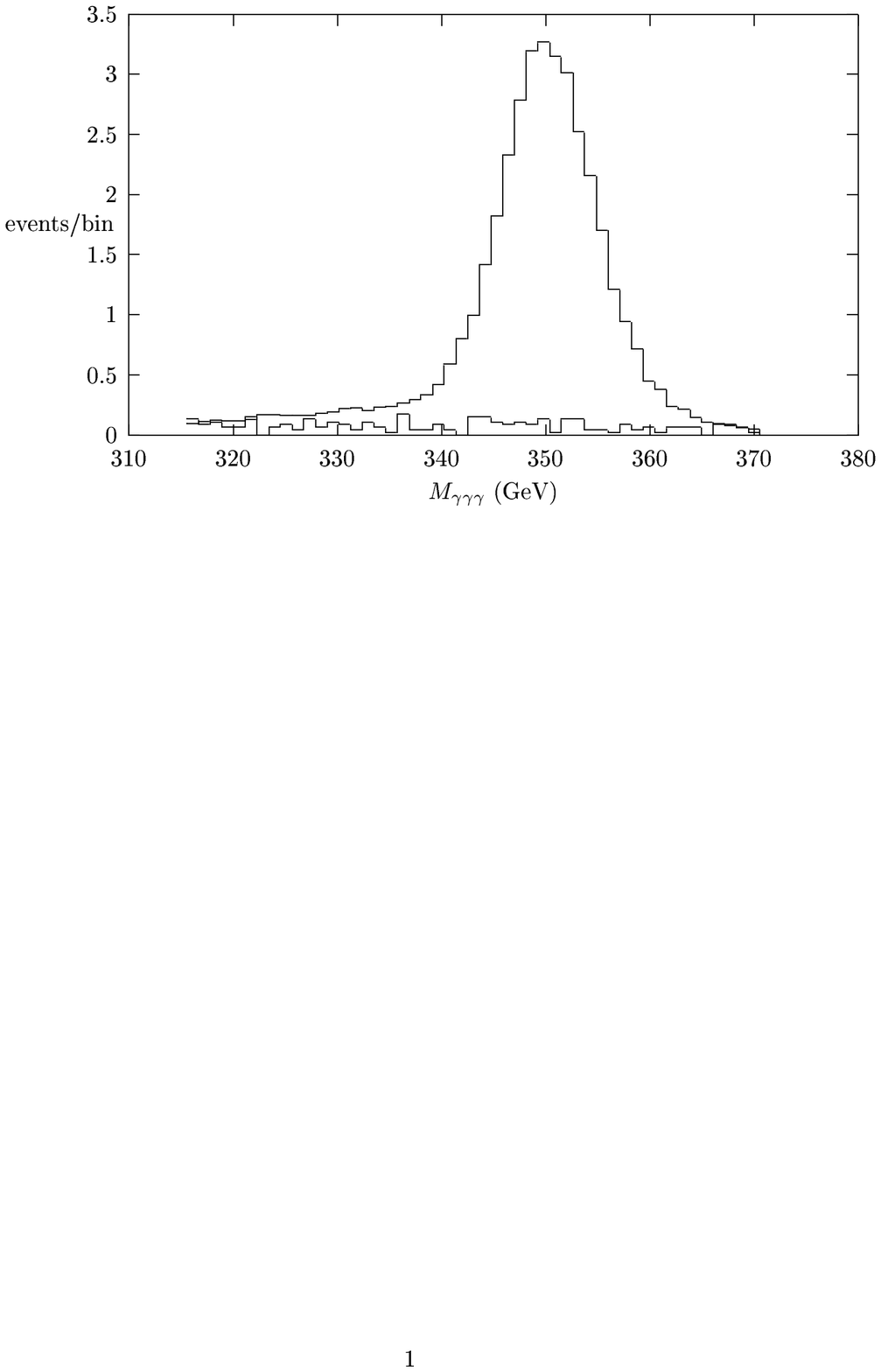}}
\vskip-20cm
\caption{3-photon invariant mass distribution for signal (upper histogram) and
background (lower histogram) for $m_{\omega_T} = m_{\rho_T} = 350$
GeV for ${\cal L} = 30$ fb$^{-1}$. The bin
size used in these histograms is $1.1$ GeV.  }
\label{3gamma}
\end{figure}

In Figures \ref{2gamma} and \ref{3gamma} we show for illustrative purposes the 
differential cross sections for signal and background as a function of   
the 2-photon and 3-photon invariant mass respectively for $m_{\omega_T} =
m_{\rho_T} =350$ GeV and $m_{\Pi_T^{(\prime)}} = 110$ GeV. In both figures 
one can clearly see a signal that stands out above the background. 
The 2-photon distribution in Fig.\ \ref{2gamma} shows a peak centered 
around the techni-pion mass (which we chose at $110$ GeV). Since 
this is a two-photon invariant mass distribution in three-photon 
events, the width of the peak does not correspond to the techni-pion
width, but it contains also the combinatoric error from the selection 
among the three photons. Indeed, for a technipion much narrower than 
the techni-vector meson, the widths in both figures are comparable.
The three-photon distribution in Fig.\ \ref{3gamma} shows a peak centered 
around the techni-vector meson mass. In this case, the width in the histogram 
reflects the resonance width together with the photon energy resolution that 
we use in the simulation. 
In addition, the distribution away from the peak is dominated by the anomaly
contribution. As it is
comparable to the background, the non-resonant contribution cannot be detected. 

In order to further suppress the background, the following cuts were used:
\begin{eqnarray}
 M_{\gamma \gamma \gamma }  
 &\in& \left[ {M_{\omega_T}   - \frac{{M_{\omega_T}  }}{{10}}, M_{\omega_T}   
 + 20 \mbox{ GeV}} 
 \right] \nonumber \\ 
 M_{\gamma \gamma }  &\in& \left[ {M_{\Pi_T}   - \frac{{M_{\Pi_T} 
 }}{{10}},M_{\Pi_T}   
 + 10\mbox{ GeV}} \right]  \nonumber \\ 
 p_{T\gamma }  &\ge& 70\mbox{ GeV} 
\end{eqnarray}

In Table I we present our results for the total number of 3-photon events for a
given techni-resonance mass and for 2 different integrated luminosities, namely
${\cal L} =$  $2$ fb$^{-1}$ and $30$ fb$^{-1}$.

\begin{table}
\begin{center}
\begin{tabular}{|c|c|c|c|c|}
\hline
$m_{\omega_T,\rho_T}$ (GeV)& $\sigma$ (fb) & Events & S/B &Significance \\
\hline
210 & 18.22 & 12 - 175 & 38.6 & 21.2 - 82.3\\
250 & 9.22  & 9  - 135 & 19.0 & 13.1 - 50.7\\
300 & 4.83  & 4.3 - 64.7 & 13.1 & 7.5 - 29.1\\
350 & 2.70  & 2.6 - 38.8 & 10.3 & 5.2 - 20.0\\
400 & 1.83  & 0.92 - 13.8& 5.0 & 2.2 - 8.4\\
450 & 1.06  & 0.8 - 12.2 & 7.8 & 2.5 - 9.8\\
500 & 1.00  & 0.2 - 3.3 & 3.5 & 0.9 - 3.4\\
550 & 0.78  & 0.2 - 3.4 & 3.5 & 0.9 - 3.5\\
600 & 0.52  & 0.06 - 0.9 & 1.6 & 0.3 - 1.2\\
\hline
\end{tabular}
\label{table}
\caption{Cross sections (before cuts), number of events (after cuts), 
signal/background ratio and
significance of the signal for ${\cal L} =$  $2$ fb$^{-1}$ and $30$ fb$^{-1}$ 
(first and second figures respectively) for different techni-resonance masses.}
\end{center}
\end{table}

\begin{figure}
\vskip-10cm
\centerline{\epsfxsize=1.5\hsize \epsffile{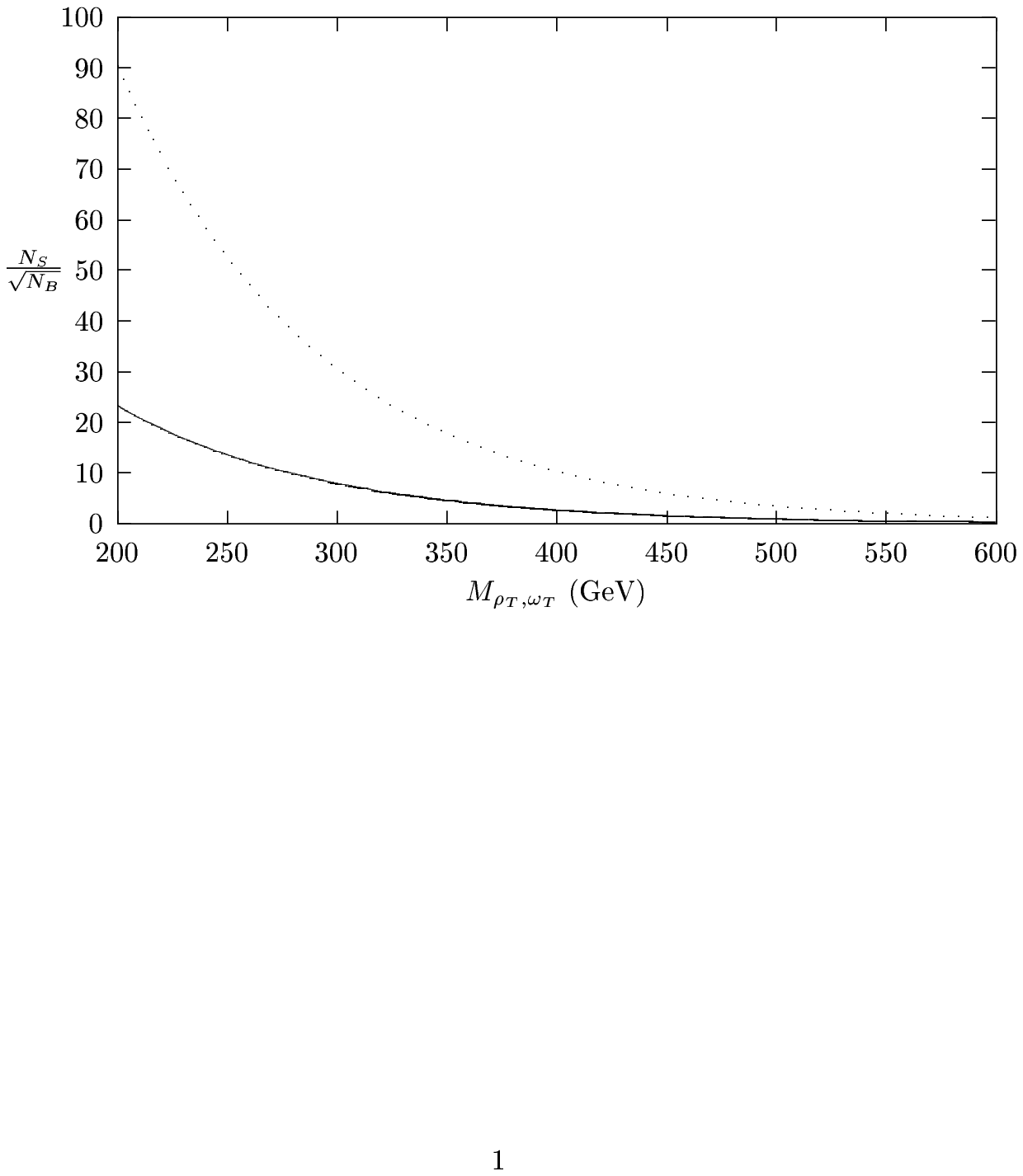}}
\vskip-12cm
\caption{Statistical significance of signal for $\cal{L}$=30 fb$^{-1}$ 
(dots) and $\cal{L}$=2 fb$^{-1}$ (solid line) as a function of the
masses of the techni-resonances. }
\label{significance}
\end{figure}

We can see that resonances up to $350$ GeV can be found at the $5 \sigma$
level even with  ${\cal L} =$  $2$ fb$^{-1}$. For an accumulated luminosity of
${\cal L} =$  $30$ fb$^{-1}$, resonances as heavy as $550$ GeV can be detected
at the $3 \sigma$ level. In Figure \ref{significance} we show the statistical 
significance of the signal as a function of the techni-resonances for the two
luminosities.

\section{Conclusions}

In this paper we have examined the triple photon production at the Tevatron as
a signature for technicolor models.
We have included both resonant and
non-resonant contributions, but the former are dominant in a hadron machine,
where the center-of-mass energy of the process is not fixed. 
The relatively low background enables one to
obtain large significance levels. We found that technicolor models can be
effectively probed in this mode and, with an accumulated luminosity of
${\cal L} =$  $30$ fb$^{-1}$, resonances as heavy as $550$ GeV can be detected
or excluded at the $3 \sigma$ level. Using this mode we can have information on
both the techni-vectors as well as the techni-pion masses from the $3-$ and $2-$
photon invariant mass distributions respectively.

\section*{Acknowledgments}

A.Z. and C.D. received partial support
from Fondecyt (Chile) grants No.~3020002 and 8000017,
respectively. R.R. would like to thank CNPq and PRONEX for partial
financial support. The authors also acknowledge support from Fundacion
Andes (Chile) and Funda\c{c}\~ao Vita (Brazil)  grant C-13680/4.  

% Journal and other miscellaneous abbreviations for references
\def \arnps#1#2#3{Ann.\ Rev.\ Nucl.\ Part.\ Sci.\ {\bf#1} (#3) #2}
\def \art{and references therein}
\def \cmts#1#2#3{Comments on Nucl.\ Part.\ Phys.\ {\bf#1} (#3) #2}
\def \cn{Collaboration}
\def \cp89{{\it CP Violation,} edited by C. Jarlskog (World Scientific,
Singapore, 1989)}
\def \econf#1#2#3{Electronic Conference Proceedings {\bf#1}, #2 (#3)}
\def \efi{Enrico Fermi Institute Report No.\ }
\def \epjc#1#2#3{Eur.\ Phys.\ J. C {\bf#1} (#3) #2}
\def \f79{{\it Proceedings of the 1979 International Symposium on Lepton and
Photon Interactions at High Energies,} Fermilab, August 23-29, 1979, ed. by
T. B. W. Kirk and H. D. I. Abarbanel (Fermi National Accelerator Laboratory,
Batavia, IL, 1979}
\def \hb87{{\it Proceeding of the 1987 International Symposium on Lepton and
Photon Interactions at High Energies,} Hamburg, 1987, ed. by W. Bartel
and R. R\"uckl (Nucl.\ Phys.\ B, Proc.\ Suppl., vol.\ 3) (North-Holland,
Amsterdam, 1988)}
\def \ib{{\it ibid.}~}
\def \ibj#1#2#3{~{\bf#1} (#3) #2}
\def \ichep72{{\it Proceedings of the XVI International Conference on High
Energy Physics}, Chicago and Batavia, Illinois, Sept. 6 -- 13, 1972,
edited by J. D. Jackson, A. Roberts, and R. Donaldson (Fermilab, Batavia,
IL, 1972)}
\def \ijmpa#1#2#3{Int.\ J.\ Mod.\ Phys.\ A {\bf#1} (#3) #2}
\def \ite{{\it et al.}}
\def \jhep#1#2#3{JHEP {\bf#1} (#3) #2}
\def \jpb#1#2#3{J.\ Phys.\ B {\bf#1} (#3) #2}
\def \jpg#1#2#3{J.\ Phys.\ G {\bf#1} (#3) #2}
\def \mpla#1#2#3{Mod.\ Phys.\ Lett.\ A {\bf#1} (#3) #2}
\def \nat#1#2#3{Nature {\bf#1} (#3) #2}
\def \nc#1#2#3{Nuovo Cim.\ {\bf#1} (#3) #2}
\def \nima#1#2#3{Nucl.\ Instr.\ Meth. A {\bf#1} (#3) #2}
\def \npb#1#2#3{Nucl.\ Phys.\ B {\bf#1} (#3) #2}
\def \npps#1#2#3{Nucl.\ Phys.\ Proc.\ Suppl.\ {\bf#1} (#3) #2}
\def \npbps#1#2#3{Nucl.\ Phys.\ B Proc.\ Suppl.\ {\bf#1} (#3) #2}
\def \PDG{Particle Data Group, D. E. Groom \ite, \epjc{15}{1}{2000}}
\def \pisma#1#2#3#4{Pis'ma Zh.\ Eksp.\ Teor.\ Fiz.\ {\bf#1} (#3) #2 [JETP
Lett.\ {\bf#1} (#3) #4]}
\def \pl#1#2#3{Phys.\ Lett.\ {\bf#1} (#3) #2}
\def \pla#1#2#3{Phys.\ Lett.\ A {\bf#1} (#3) #2}
\def \plb#1#2#3{Phys.\ Lett.\ B {\bf#1} (#3) #2}
\def \pr#1#2#3{Phys.\ Rev.\ {\bf#1} (#3) #2}
\def \prc#1#2#3{Phys.\ Rev.\ C {\bf#1} (#3) #2}
\def \prd#1#2#3{Phys.\ Rev.\ D {\bf#1} (#3) #2}
\def \prl#1#2#3{Phys.\ Rev.\ Lett.\ {\bf#1} (#3) #2}
\def \prp#1#2#3{Phys.\ Rep.\ {\bf#1} (#3) #2}
\def \ptp#1#2#3{Prog.\ Theor.\ Phys.\ {\bf#1} (#3) #2}
\def \ppn#1#2#3{Prog.\ Part.\ Nucl.\ Phys.\ {\bf#1} (#3) #2}
\def \rmp#1#2#3{Rev.\ Mod.\ Phys.\ {\bf#1} (#3) #2}
\def \rp#1{~~~~~\ldots\ldots{\rm rp~}{#1}~~~~~}
\def \si90{25th International Conference on High Energy Physics, Singapore,
Aug. 2-8, 1990}
\def \zpc#1#2#3{Zeit.\ Phys.\ C {\bf#1} (#3) #2}
\def \zpd#1#2#3{Zeit.\ Phys.\ D {\bf#1} (#3) #2}

\newpage

\section*{Table captions}
\vskip1cm

Table I:
Cross sections, number of events, signal/background ratio and
significance of the signal for ${\cal L} =$  $2$ fb$^{-1}$ and $30$ fb$^{-1}$ 
(first and second figures respectively) for different techni-resonance masses.

\section*{Figure captions}
\vskip1cm

\hskip0.6cm Figure 1: 
Triangle anomaly giving the continuum contribution to the 
$p\bar{p} \to \Pi_T \gamma$ process.

Figure 2:
Feynman diagrams for the process
$e^+e^-\to \tau^+\tau^-  \nu_\mu\bar\nu_\mu$.

Figure 3:
2-photon invariant mass distribution for signal (upper histogram) and
background (lower histogram) for $m_{\omega_T} = m_{\rho_T} = 350$
GeV and $m_{\Pi_T^{(\prime)}} = 110$ GeV for ${\cal L} = 30$ fb$^{-1}$. The bin
size used in  these histograms is $0.43$ GeV. 

Figure 4:
3-photon invariant mass distribution for signal (upper histogram) and
background (lower histogram) for $m_{\omega_T} = m_{\rho_T} = 350$
GeV for ${\cal L} = 30$ fb$^{-1}$. The bin
size used in these histograms is $1.1$ GeV. 

Figure 5:
Statistical significance of signal for $\cal{L}$=30 fb$^{-1}$ 
(dots) and $\cal{L}$=2 fb$^{-1}$ (solid line) as a function of the
masses of the techni-resonances.

\newpage

%\begin{figure}
%\centerline{\epsfxsize=1\hsize \epsffile{2gamma.ps}
%\hskip-7cm \epsfxsize=1\hsize \epsffile{3gamma.ps}}
%\vskip-2cm
%\caption{}
%\label{fig1}
%\end{figure}

\end{document}